\documentclass[aps,prx,twocolumn,longbibliography,floatfix,10pt,superscriptaddress]{revtex4-2}
\usepackage[utf8]{inputenc}
\usepackage{mathrsfs}
\usepackage{natbib}
\usepackage{graphicx}
\usepackage{latexsym}
\usepackage{amssymb}
\usepackage{amsmath}
\usepackage{amsfonts}
\usepackage{gensymb}
\usepackage{bm}
\usepackage{hyperref}
\usepackage{braket}
\usepackage{physics}
\usepackage{bbold}
\usepackage[caption=false]{subfig}
\usepackage[dvipsnames]{xcolor}
\usepackage[normalem]{ulem}
\usepackage{siunitx}
\usepackage{mhchem}
\usepackage{multirow}
\usepackage{soul}

\usepackage{slashed}

\definecolor{myforestgreen}{RGB}{34,139,34}

\newcommand{\PRLsec}[1]{\emph{#1---}}

\newcommand{\supplementarysection}{%
  \setcounter{Fig.}{0}
  \let\oldtheFig.\theFig.
  \renewcommand{\theFig.}{S\oldtheFig.}
  \setcounter{section}{0}
  \let\oldthesection\thesection
  \renewcommand{\thesection}{S\oldthesection}
  \setcounter{equation}{0}
  \let\oldtheequation\theequation
  \renewcommand{\theequation}{S\oldtheequation}
  \setcounter{table}{0}
  \let\oldthetable\thetable
  \renewcommand{\thetable}{S\oldthetable}
}

\newenvironment{dfn}{{\vspace*{1ex} \noindent \bf Definition }}{\vspace*{1ex}}

	\newcommand{\beq}{\begin{eqnarray}}
	\newcommand{\eeq}{\end{eqnarray}}
	\newcommand{\bea}{\begin{eqnarray}\begin{aligned}}
	\newcommand{\eea}{\end{aligned}\end{eqnarray}}

\begin{document}

\title{Expressibility of neural quantum states: a Walsh-complexity perspective}

\author{Taige Wang}
\affiliation{Department of Physics, Massachusetts Institute of Technology, Cambridge, MA 02139, USA \looseness=-2}
\affiliation{Department of Physics, Harvard University, Cambridge, MA 02138, USA \looseness=-2}

\begin{abstract}
Neural quantum states are powerful variational wavefunctions, but it remains unclear which many-body states can be represented efficiently by modern additive architectures. We introduce \emph{Walsh complexity}, a basis-dependent measure of how broadly a wavefunction is spread over parity patterns. States with an almost uniform Walsh spectrum require exponentially large Walsh complexity from any good approximant. We show that shallow additive feed-forward networks cannot generate such complexity in a controlled regime, e.g. fixed-degree polynomial activations with subexponential parameter scaling. As a concrete example, we construct a simple dimerized state prepared by a single layer of disjoint controlled-$Z$ gates. Although it has only short-range entanglement and a simple tensor-network description, its Walsh complexity is maximal. Full-cube fits across system size and depth are consistent with the complexity bound: for polynomial activations, successful fitting appears only once depth reaches a logarithmic scale in $N$, whereas activation saturation in $\tanh$ produces a sharp threshold-like jump already at depth $3$. Walsh complexity therefore provides an expressibility axis complementary to entanglement and clarifies when depth becomes an essential resource for additive neural quantum states.
\end{abstract}

\maketitle

Neural quantum states (NQS) provide flexible variational wavefunctions across many-body physics~\cite{CarleoTroyer2017,DengLiDasSarma2017,SharirShashuaCarleo2022}, yet a quantitative theory of efficient representability remains incomplete: with only $\mathrm{poly}(N)$ trainable parameters, which $N$-body states admit faithful representation by NQS?

Restricted Boltzmann machines (RBMs) furnish the clearest benchmark.
Their correlator-product form gives exact $\mathrm{poly}(N)$ descriptions of broad stabilizer and graph-state families~\cite{GaoDuan2017,Jia2019,LuGaoDuan2019}, but also explicit limitations: the GWD family, obtained from a two-dimensional cluster state by one layer of local unitaries, has no efficient RBM representation~\cite{GaoDuan2017}.
More generally, efficiently contractible tensor-network states such as MPS can be embedded into multiplicative NQS with depth $O(\log N)$~\cite{SharirShashuaCarleo2022}.

Modern NQS, however, increasingly use additive parameterizations~\cite{SpragueCzischek2024}.
We distinguish \emph{additive} and \emph{multiplicative} coefficient models by how the coefficient $\psi(\sigma)$ is built.
In additive models the readout is composed along the computation path, whereas in multiplicative models it becomes the product of scalar factors of the entire network.
Feed-forward or transformer backbones used as direct coefficient models are additive in this sense, whereas RBMs and autoregressive factorizations are multiplicative at the coefficient level.
Log-space rewritings do not remove this multiplicative resource at the level of coefficient construction~\footnote{Near coefficient zeros the logarithm is singular and can worsen numerical stability during training, so the rewriting is not innocuous from the optimization viewpoint either~\cite{SharirShashuaCarleo2022,GoodfellowBengioCourville2016}.}.

For additive architectures without built-in geometry, real-space entanglement is often a poor proxy for expressibility~\cite{Paul2025EntanglementBound}: even shallow NQS can already support volume-law entanglement~\cite{DengLiDasSarma2017}.
We therefore fix the computational basis and rescale coefficients as $f(\sigma)\equiv 2^{N/2}\psi(\sigma)$, turning the normalized wavefunction into a function on the Boolean cube and analyzing it in the Walsh--Hadamard basis~\cite{ODonnellBook}.
We define
\begin{equation}
\|f\|_W\equiv\sum_{S\subseteq[N]}|\widehat f(S)|,
\label{eq:walsh_norms_intro}
\end{equation}
where $\widehat f(S)$ are Walsh coefficients.
$\|f\|_W$ measures how broadly the state is spread over parity patterns in the conjugate basis.
Two simple inequalities then organize the expressibility problem,
\begin{equation}
|\langle f,g\rangle|\le \|\widehat f\|_{\infty}\,\|g\|_W,
\qquad
\|fg\|_W\le \|f\|_W\,\|g\|_W.
\label{eq:overlap_submult_intro}
\end{equation}
The first turns $\|g\|_W$ into a necessary approximation resource, and the second shows why the complexity grows easily in multiplicative models.

A central benchmark appears once this notion is in place.
We construct a dimerized state prepared by one layer of disjoint controlled-$Z$ gates.
It has only short-range dimer entanglement and an exact bond-dimension-$2$ MPS description, yet its coefficient pattern is a quadratic bent function with perfectly flat Walsh spectrum, saturating $\|f\|_W=2^{N/2}$.
This gives a minimal many-body example in which entanglement and tensor-network simplicity are both misleading proxies for additive NQS expressibility.

We formulate a Walsh-spectral expressibility theory in coefficient space and prove our main theorem for the real scalar function represented by a canonical additive feed-forward architecture.
In the tame regime—for example, fixed-degree polynomial activations with subexponential parameter scaling—constant-depth additive networks satisfy $\|g\|_W=\exp(o(N))$.
For the equal-weight Walsh-flat targets studied here, including the dimer benchmark, this rules out $O(1)$ overlap.
Full-cube fits across $N$ and $D$ at linear width $w=2N$ are consistent with this obstruction: in the tame polynomial regime, success onsets only once depth reaches a logarithmic scale in $N$, whereas bounded activations such as $\tanh$ display a sharp threshold-like transition already near $D=3$.

For bounded activations such as $\tanh$, the picture changes once preactivations enter saturation.
The network then approximates threshold gates, and for equal-weight Boolean readouts the problem moves into the regime of constant-depth threshold circuits ($TC^0$).
For general $TC^0$, explicit superpolynomial lower bounds are notoriously scarce~\cite{ViolaThesis2006,AllenderFSTTCS99,ChenLyu2021,Parham2025MagicHierarchy,RazborovRudich1997}, which explains why explicit lower bounds become difficult in the threshold regime.
One message of the present work is therefore twofold: in the tame additive regime one can prove sharp Walsh complexity ceilings, while beyond that regime NQS can appear extraordinarily expressive in practice.

\PRLsec{Two canonical examples and the Walsh spectrum}
\label{sec:canonical}
We consider $N$ qubits in the computational ($Z$) basis $\sigma=(\sigma_1,\dots,\sigma_N)\in\{\pm1\}^N$, where $\sigma_i=\pm1$ is the eigenvalue of $Z_i$.
A wavefunction is a function $\psi:\{\pm1\}^N\to\mathbb{C}$, and we write $f(\sigma)\equiv 2^{N/2}\psi(\sigma)$.

\begin{figure}[t]
\includegraphics[width=\columnwidth]{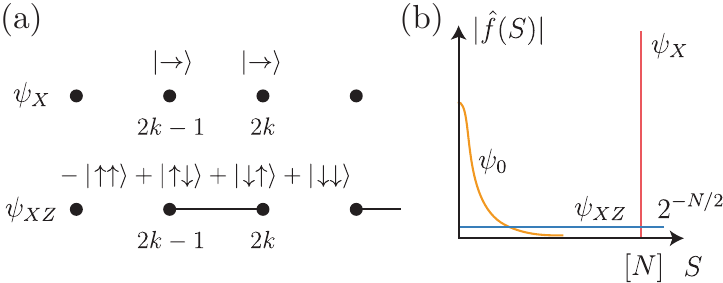}
\caption{\textbf{Two canonical examples and their Walsh spectra.}
(a) Benchmark states $\psi_X$ and $\psi_{XZ}$.
(b) Schematic spectra: a single spike for $\psi_X$, a flat spectrum for $\psi_{XZ}$, and a generic few-body profile with weight concentrated at small $|S|$.}
\label{fig:canonical}
\end{figure}

For any subset $S\subseteq[N]$, define the parity character $\chi_S(\sigma)=\prod_{i\in S}\sigma_i$.
For any $h:\{\pm1\}^N\to\mathbb{C}$,
\begin{equation}
\widehat h(S)=2^{-N}\sum_{\sigma}h(\sigma)\chi_S(\sigma),
\qquad
h(\sigma)=\sum_{S\subseteq[N]}\widehat h(S)\chi_S(\sigma).
\label{eq:walsh}
\end{equation}
For a normalized wavefunction, $\widehat f(S)$ is the $X$-basis amplitude of the product state $\ket{S}_X\equiv\bigotimes_{i\in S}\ket{-}_i\bigotimes_{i\notin S}\ket{+}_i$, so with $p_S\equiv |\widehat f(S)|^2$, one has \begin{equation} \|f\|_W=\sum_{S\subseteq[N]}\sqrt{p_S}, \qquad 2\log_2 \|f\|_W=H_{1/2}(p), \label{eq:renyi_half} \end{equation} so Walsh complexity is the R\'enyi-$\tfrac12$ entropy of the $X$-basis outcome distribution in exponential form. Parseval gives
\begin{equation}
    1\le \|f\|_W\le 2^{N/2},
\end{equation}
with equality at the upper end if the spectrum is flat.

Walsh complexity already constrains approximation before any architecture is specified.
Since $\langle f,g\rangle=2^{-N}\sum_\sigma f(\sigma)^*g(\sigma)=\sum_{S}\widehat f(S)^*\widehat g(S)$, one has
\begin{equation}
|\langle f,g\rangle|
\le
\|\widehat f\|_{\infty}\,\|g\|_W,
\qquad
\|\widehat f\|_{\infty}\equiv \max_{S\subseteq[N]}|\widehat f(S)|.
\label{eq:overlap_general}
\end{equation}
Thus $\|g\|_W$ is not merely a diagnostic but a necessary approximation resource.
If the target has $\|\widehat f\|_{\infty}=\exp(-\Omega(N))$ while the ansatz can realize only $\|g\|_W=\exp(o(N))$, then the overlap remains exponentially small.
This does not contradict universal approximation: it identifies a resource threshold that must be crossed before approximation can begin.

As a minimal reference point, the $x$-polarized product state $\ket{-}^{\otimes N}$ has $f_X(\sigma)=\chi_{[N]}(\sigma)$, hence \begin{equation} \widehat f_X(S)=\delta_{S,[N]}, \qquad \|f_X\|_W=1. \label{eq:fX_spectrum} \end{equation}
All spectral weight sits in a single Walsh mode.

Our main example is the ground state of the dimerized frustration-free commuting-Pauli Hamiltonian
\begin{equation} \label{eq:HXZ} \begin{gathered} H_{XZ}=-\sum_{k=1}^{N/2}\Big(X_{2k-1}Z_{2k}+Z_{2k-1}X_{2k}\Big),\\ \ket{\psi_{XZ}}=\bigotimes_{k=1}^{N/2}\ket{\psi_{2k-1,2k}}, \end{gathered} \end{equation} where \begin{equation} \ket{\psi_{2k-1,2k}}=\frac12\Big(\ket{\uparrow\uparrow}+\ket{\uparrow\downarrow}+\ket{\downarrow\uparrow}-\ket{\downarrow\downarrow}\Big) = CZ_{12}\ket{+}\ket{+}, \label{eq:psi12} \end{equation} 
i.e. a two-vertex graph state~\cite{HeinEisertBriegel2004}.
Thus $\ket{\psi_{XZ}}$ is prepared by a single layer of disjoint controlled-$Z$ gates acting on $\ket{+}^{\otimes N}$.
It has only dimer entanglement and an exact bond-dimension-$2$ MPS description, yet its coefficients are maximally delocalized in Walsh space.
The same state is therefore expressible for multiplicative NQS: by Ref.~\cite{SharirShashuaCarleo2022}, it can also be realized as a multiplicative NQS with depth $O(\log N)$.

For one dimer, the coefficient pattern equals $-1$ only at $(\sigma_1,\sigma_2)=(-1,-1)$ and $+1$ otherwise, so $|\widehat f_{12}(S)|=1/2$ for all $S\subseteq\{1,2\}$.
Because the full coefficient pattern factorizes over dimers, the Walsh coefficients factorize as well:
\begin{equation}
|\widehat f_{XZ}(S)|=2^{-N/2}
\quad
\text{for all }S\subseteq[N],
\qquad
\|f_{XZ}\|_W=2^{N/2}.
\label{eq:flat}
\end{equation}
In $\{0,1\}$ variables $x_i=(1-\sigma_i)/2$, 
\begin{equation} f_{XZ}(x)=(-1)^{\sum_{k=1}^{N/2} x_{2k-1}x_{2k}}, \label{eq:ip_mod_2} \end{equation}
is a canonical quadratic \emph{bent} function called inner-product mod $2$ (IP$_2$) with flat Walsh spectrum~\cite{Rothaus1976,CanteautCharpin2003,ODonnellBook}. Thus $\psi_{XZ}$ is a minimal many-body example with limited entanglement but maximal Walsh complexity.

A generic bounded coefficient pattern already has near-flat Walsh statistics. If $\{f(\sigma)\}$ are independent with mean zero and variance $\langle |f(\sigma)|^2\rangle=\varsigma^2$, then for any nonempty $S$ one has $\langle |\widehat f(S)|^2\rangle=\varsigma^2 2^{-N}$, so typically $|\widehat f(S)|\sim \varsigma 2^{-N/2}$ and hence $\|f\|_W\sim \varsigma 2^{N/2}$ up to constants. By contrast, coefficient patterns generated by few-body structure in the chosen basis have weight concentrated on small subsets with a decaying tail toward large $|S|$, as sketched in Fig.~\ref{fig:canonical}(b). We use $f_{XZ}$ below as a stringent and analytically tractable Walsh-hard target.

\PRLsec{Expressibility in the tame regime}
\label{sec:expressibility}
The overlap bound in Eq.~\eqref{eq:overlap_general} reduces NQS expressibility to a concrete question: how much Walsh complexity can a shallow feed-forward network generate?

\begin{figure}[t]
\includegraphics[width=0.8\columnwidth]{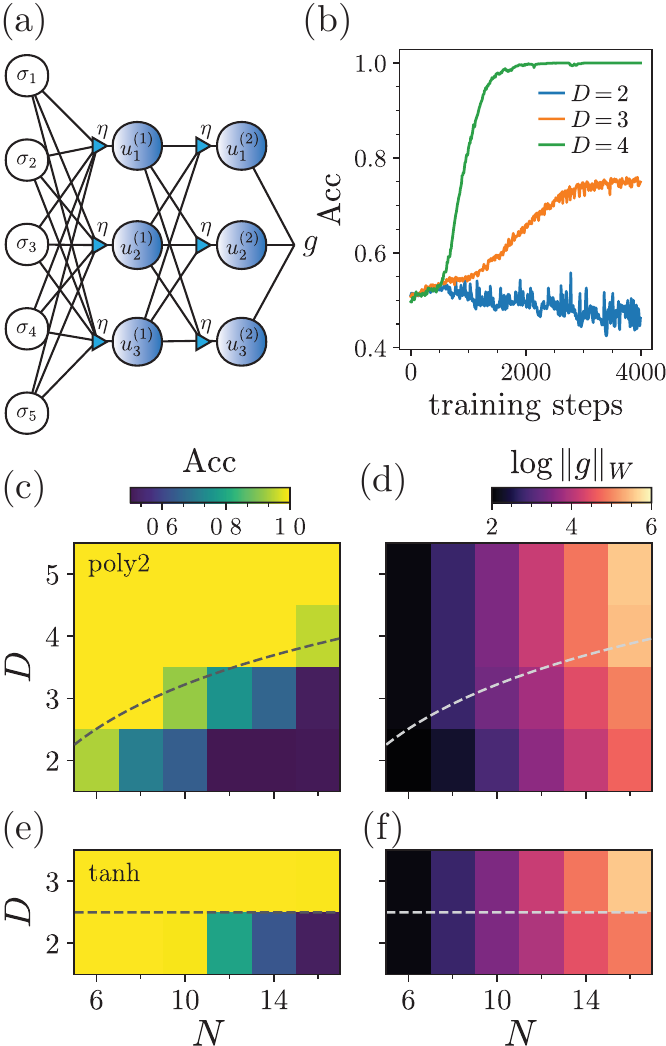}
\caption{\textbf{Exact fitting as an expressibility test.}
Fitting the bent dimer target $f_{XZ}(\sigma)$ on the full hypercube with hidden width $w=2N$.
(a) Additive feed-forward scalar network.
(b) Representative full-cube accuracy during training at $N=12$ for the degree-$2$ polynomial activation.
(c,e) Final full-cube accuracy of the Boolean readout $\tilde g_\theta(\sigma)=\mathrm{sign}(g_\theta(\sigma))$.
(d,f) Corresponding Walsh complexity $\log\|\tilde g_\theta\|_W$.
(c,d) Degree-$2$ polynomial activation. (e,f) $\tanh$ activation.
The dashed curve in (c,d) marks the predicted depth scale $D\approx \log N$, and the dashed line in (e,f) marks the threshold depth $D=3$.}
\label{fig:expressibility}
\end{figure}

We therefore consider a real-valued additive feed-forward scalar model of depth $D$ and hidden width $w$ with Boolean input $\sigma\in\{\pm1\}^N$.
Here $D$ counts the $D-1$ hidden layers together with the input layer; the hidden layers are indexed by $\ell=2,\dots,D$, each with $w$ neurons.
Absorbing the final linear readout into a formal scalar output layer $\ell=D+1$, the network is shown in Fig.~\ref{fig:expressibility}(a).
\begin{equation} \begin{gathered} u_i^{(1)}(\sigma)=\sigma_i, \qquad i=1,\dots,N,\\ z_j^{(\ell)}(\sigma) =\sum_i W_{ji}^{(\ell)}\,u_i^{(\ell-1)}(\sigma)+b_j^{(\ell)}, \qquad \ell=2,\dots,D,\\ u_j^{(\ell)}(\sigma)=\eta\bigl(z_j^{(\ell)}(\sigma)\bigr), \qquad \eta:\mathbb{R}\to\mathbb{R}\\ g(\sigma) =\sum_{j=1}^{w} W_{1j}^{(D+1)}\,u_j^{(D)}(\sigma)+b_1^{(D+1)}. \end{gathered} \label{eq:mlp} \end{equation}
The activation $\eta$ is applied elementwise.
At the coefficient level this is additive: the output is built by repeated composition of affine maps and scalar nonlinearities. The additive theorem below, and the Barron-type statement later on, are written for real scalar functions. Complex-valued wavefunctions may be handled by decomposing into real and imaginary parts $f=f_R+i f_I$ with $f_R,f_I:\{\pm1\}^N\to\mathbb{R}$, and the Walsh complexity obeys
$\|f\|_W\le \|f_R\|_W+\|f_I\|_W$.

Our bound propagates Walsh mass through the computational graph using the absolute Taylor majorant of the activation.
Writing $\eta(x)=\sum_{r\ge0}a_r x^r$, define $\widetilde\eta(R)\equiv\sum_{r\ge0}|a_r|R^r$.
Using subadditivity, submultiplicativity, and $\|\eta\circ h\|_W\le \widetilde\eta(\|h\|_W)$ whenever the right-hand side is finite, we obtain (see Appendix for proof):

\textbf{Theorem (tame-majorant bound).}
Assume $\eta$ is analytic with absolute majorant $\widetilde\eta$.
Define
\begin{equation}
\mathcal{W}\equiv\max_{\ell,j}\sum_i |W_{ji}^{(\ell)}|,
\qquad
B\equiv\max_{\ell,j}|b_j^{(\ell)}|,
\label{eq:kappa_def}
\end{equation}
and let $R_1\equiv1$, $R_\ell\equiv \widetilde\eta(B+\mathcal{W}R_{\ell-1})$ for $\ell=2,\dots,D$.
Then
\begin{equation}
\|g\|_W\le B+\mathcal{W}R_D.
\label{eq:g0_bound}
\end{equation}
In the fully connected width-$w$ case with entrywise bounds $|W_{ji}^{(\ell)}|\le s$, one may take $\mathcal{W}\le s\max(N,w)$.

The recursion is informative only while it remains tame, meaning that $\widetilde\eta$ stays finite on the generated range and grows subexponentially in $N$.
For degree-$p$ polynomial activations one finds
\begin{equation}
\|g\|_W\lesssim K^{O(p^{D-1})},
\qquad
K\equiv 2+\mathcal{W}+B.
\label{eq:poly_growth}
\end{equation}

\textbf{Corollary.}
For degree-$p$ polynomial activations with $K=\mathrm{poly}(N)$ and depth $D\le (1-\varepsilon)\log_p N$, additive networks satisfy $\|g\|_W=\exp(o(N))$.

Combined with Eq.~\eqref{eq:overlap_general}, this immediately excludes $O(1)$ overlap with Walsh-flat targets such as $f_{XZ}$ in the tame regime.
This is a finite-resource obstruction, not a contradiction to universal approximation.

Two scope conditions matter.
First, the activation must remain tame on the relevant preactivation range.
Entire nonpolynomial activations such as $e^x$, $\sin x$, and $\cos x$ have finite majorants for all $R$, but these still grow exponentially once preactivations become extensive.
Bounded analytic activations such as $\tanh$ or the logistic sigmoid are even less tame from the majorant viewpoint: $\widetilde{\tanh}(R)=\tan R$ diverges already at $R=\pi/2$.
Thus Eq.~\eqref{eq:g0_bound} is informative only in a small preactivation regime.

Second, the preactivation parameters $\mathcal{W}$ and $B$ must themselves remain subexponential.
Otherwise Eq.~\eqref{eq:g0_bound} is already exponential and gives no useful obstruction.
The theorem is therefore most useful for targets with $\|f\|_\infty\le \exp(o(N))$, for example equal-weight states such as $\psi_X$ and $\psi_{XZ}$, or for the phase angle of $f$ in a fixed branch.
The latter only constrains exact representability of the phase channel, not the overlap bound above.
In the Boolean case $f(\sigma)\in\{\pm1\}$, however, the coefficient pattern is affinely equivalent to the phase angle, so the distinction disappears.

\PRLsec{Exact fitting and threshold behavior}
\label{sec:numerics}
We now test the obstruction directly on $f_{XZ}$ by fitting the full Boolean cube across system size $N$ and depth $D$, with hidden width fixed to $w=2N$.
The theorem controls the pre-threshold logit $g_\theta(\sigma)$.
Numerically, however, directly optimizing a Boolean output is ill-conditioned, so we cast the task as binary classification and train the logits using
\begin{equation}
\mathcal{L}(\theta)=\Big\langle \log\big(1+\exp[-f_{XZ}(\sigma)g_\theta(\sigma)]\big)\Big\rangle,
\label{eq:classification_loss}
\end{equation}
while evaluating the induced Boolean readout $\tilde g_\theta(\sigma)\equiv \mathrm{sign}(g_\theta(\sigma))$.
Operationally, this appends a final threshold gate to the additive logit model and therefore probes a hypothesis class larger than that covered by the theorem.
For each $N$, the training set is the full cube $\{\pm1\}^N$. At each gradient step we sample a random minibatch of size $512$ from this full set.
Unless noted otherwise, we train for $12000$ gradient steps and report the final full-cube accuracy and Walsh complexity of the readout.
Fig.~\ref{fig:expressibility}(b) shows a representative $N=12$ training trace, while panels~(c)--(f) of Fig.~\ref{fig:expressibility} summarize the final full-cube metrics across the $(N,D)$ sweep.

For each trained model we evaluate the Boolean readout on the full cube and record the full-cube accuracy
\begin{equation}
\mathrm{Acc}
=
\frac{1+\langle f_{XZ},\tilde g_\theta\rangle}{2}
\le
\frac12+2^{-N/2-1}\,\|\tilde g_\theta\|_W.
\label{eq:accuracy_def}
\end{equation}
as well as its Walsh complexity $\|\tilde g_\theta\|_W$, where we have used $f_{XZ}(\sigma) \in \{ \pm 1 \}$.
Therefore, to obtain $O(1)$ accuracy above chance, the readout itself must acquire Walsh complexity of order $2^{N/2}$.
The Walsh complexity is thus a direct diagnostic of whether the network is generating a useful approximant.

For the degree-$2$ polynomial activation, the representative $N=12$ training curves already show the depth dependence clearly: $D=2$ stays near chance, $D=3$ improves only partially, and $D=4$ reaches exact fitting [Fig.~\ref{fig:expressibility}(b)].
In the full $N$--$D$ sweep in Fig.~\ref{fig:expressibility}(c,d), the final accuracy frontier tracks the dashed curve $D\approx \log N$, and $\log\|\tilde g_\theta\|_W$ grows in tandem, approaching the required $O(N)$ scale only beyond that frontier.
This is exactly the pattern suggested by the tame-majorant bound: with only linear width $w=2N$, successful fitting in the additive polynomial regime requires increasing compositional depth, roughly on the predicted logarithmic scale. Moreover, Fig.~\ref{fig:expressibility}(c,d) shows that even a modest shortfall of $\log\|\tilde g_\theta\|_W$ from the required $O(N)$ scaling is accompanied by a sharp drop in fitting accuracy.

Classical two-layer approximation theory gives a complementary width-only statement.
Define the first weighted Walsh moment $\|f\|_B\equiv\sum_{S\subseteq[N]}|S|\,|\widehat f(S)|$.
For sigmoidal activations and two-layer additive networks with $M$ hidden units, one has~\cite{Barron1993,Bach2017}
\begin{equation}
2^{-N}\sum_{\sigma}(f(\sigma)-g_M(\sigma))^2\lesssim \frac{\|f\|_B^2}{M}.
\end{equation}
For the flat-spectrum target $f_{XZ}$, $\|f\|_B\sim N2^{N/2}$, so these constructive guarantees require exponentially many hidden units to obtain small mean-square error.
This is the width-only counterpart to our depth-based obstruction, and the poor performance of the $D=2$ runs in Fig.~\ref{fig:expressibility}(c,d) despite the linear scaling $w=2N$ is consistent with it.

For bounded activations such as $\tanh$, the numerics show a much sharper transition.
In Fig.~\ref{fig:expressibility}(e,f), depth $D=2$ fails once $N$ becomes moderately large, whereas $D=3$ already yields exact fitting together with a rapid rise of $\log\|\tilde g_\theta\|_W$ across the full range shown.
This matches an explicit depth-$3$ threshold construction.
Let $m\equiv N/2$.
Since $f_{XZ}(x)=(-1)^{\sum_{k=1}^{m}x_{2k-1}x_{2k}}$ is parity of pairwise ANDs, it admits
\begin{equation}
f_{XZ}(\sigma)
=
2\sum_{t=0}^{m}(-1)^t
\Theta\Big(
\sum_{k=1}^{m}\Theta(-\sigma_{2k-1}-\sigma_{2k}-1)-t
\Big)-1,
\label{eq:thr_depth3_compact}
\end{equation}
where $\Theta(z)$ is the step function.
Replacing each $\Theta$ by a high-gain $\tanh$ yields a depth-$3$ additive NQS.
At depth $2$, by contrast, strong lower bounds for $\mathrm{IP}_2$ are known for several restricted threshold classes~\cite{HajnalMaassPudlakSzegedyTuran1993,Amano2020IP2}.
The $D=2$ to $D=3$ jump in the numerics is therefore not accidental: it reflects a genuine change in the underlying circuit-complexity regime.

More broadly, once bounded activations are driven into saturation, an additive network with Boolean readout $g(\sigma)\in\{\pm1\}$ effectively computes by stacking many threshold decisions in a few layers.
This places it in the same qualitative regime as $TC^0$, the class of polynomial-size, constant-depth threshold circuits~\cite{AllenderFSTTCS99,ViolaThesis2006}.
Counting arguments imply that many Boolean functions still lie outside this class, but explicit superpolynomial lower bounds are notoriously hard to prove~\cite{AllenderFSTTCS99,ViolaThesis2006,ChenLyu2021,Parham2025MagicHierarchy}.
This difficulty is not merely historical.
Natural-proofs considerations~\cite{RazborovRudich1997} suggest that broad lower-bound methods based on generic statistical signatures are unlikely to succeed in the presence of sufficiently strong pseudorandomness.
At the same time, results placing nontrivial pseudorandom primitives in $TC^0$ under standard assumptions~\cite{KrauseLucks2001} indicate that even this shallow threshold-circuit regime can already realize functions that look structureless to such generic tests.
For our purposes, the message is that once additive NQS leave the tame regime and enter threshold-like computation, one should no longer expect a simple general obstruction comparable to the Walsh-spectral ceiling proved above.
This does not mean that arbitrary targets are efficiently representable, but it does clarify why saturated NQS can appear dramatically more expressive in practice: in the corresponding circuit regime, explicitly identifying states outside the representable class is exceptionally difficult.

\PRLsec{Discussion}
Walsh complexity is a basis-resolved notion of many-body structure complementary to entanglement.
For architectures without built-in geometric locality it provides a sharp expressibility axis: the dimer bent state is Walsh-maximal despite being locally simple, short-range entangled, and MPS-exact.

The same framework also clarifies why additive and multiplicative NQS obey different representational heuristics.
For multiplicative models, complexity can accumulate through factor count, as already visible from $\|fg\|_W\le \|f\|_W\|g\|_W$.
A canonical example is the RBM,
\begin{equation}
\psi_{\rm RBM}(\sigma)=\exp(a^\top \sigma)\prod_{j=1}^{M}2\cosh\Big(b_j+W_j^\top \sigma\Big),
\label{eq:rbm_factorization}
\end{equation}
written explicitly as a product of $M$ factors.
Autoregressive NQS and contraction-based tensor-network constructions~\cite{SharirShashuaCarleo2022} exploit the same multiplicative resource.

Our results also separate two analytically distinct regimes for additive models.
In the tame regime one can propagate Walsh complexity through the computational graph and obtain explicit subexponential ceilings.
Beyond that regime, once bounded activations saturate and threshold computation becomes available, the lower-bound problem begins to resemble the general $TC^0$ problem.
The appearance of natural-proofs barriers and pseudorandom primitives in that setting is therefore part of the explanation for why modern NQS can look extraordinarily expressive once they move beyond the tame regime.

Finally, expressibility is distinct from trainability: even representable states may be hard to learn variationally. Understanding when Walsh-spectral expressibility translates into scalable optimization remains an important open problem.

\textbf{Acknowledgments.} We especially thank Liang Fu and Xuangui Huang for valuable discussions and reading of the manuscript. We also thank Nisarga Paul and Michael Zaletel for helpful discussions. We thank NEQT 2025, during which this project was initiated. T.W. is grateful for the support by the Harvard Quantum Initiative Fellowship and the Simons Collaboration on Ultra-Quantum Matter, which is a grant from the Simons Foundation (Grant No. 651440).

\appendix


\section{Lemma: Basic calculus for the Walsh $\ell_1$ norm}

\textbf{Lemma.} \textit{(Properties of $\|\,\cdot\,\|_W$)}
Let $f,g:\{\pm1\}^N\to\mathbb{C}$ and define $\|f\|_W\equiv\sum_{S\subseteq[N]}|\widehat f(S)|$, where
$\widehat f(S)=2^{-N}\sum_{\sigma} f(\sigma)\chi_S(\sigma)$ and $\chi_S(\sigma)=\prod_{i\in S}\sigma_i$.
\begin{itemize}
\item[(1)] \textit{(Subadditivity)} $\|f+g \|_{W}\le \|f \|_{W}+\|g \|_{W}$.
\item[(2)] \textit{(Products and powers)}
\begin{equation}
\|fg \|_{W}\le \|f \|_{W}\|g \|_{W},
\qquad
\|f^{r} \|_{W}\le \|f \|_{W}^{r}\ \ (r\in\mathbb{N}).
\label{eq:product0_powers0}
\end{equation}
\item[(3)] \textit{(Affine functions)} For $z(\sigma)=b+\sum_i w_i\sigma_i$,
\begin{equation}
\|z \|_{W}=|b|+\sum_i|w_i|.
\label{eq:affine0}
\end{equation}
\item[(4)] \textit{(Analytic composition and absolute Taylor majorant)}
Suppose $\eta$ is analytic at the origin with Taylor series $\eta(x)=\sum_{r=0}^{\infty}a_r x^r$.
Define the absolute Taylor majorant
\begin{equation}
\widetilde\eta(R)\equiv\sum_{r=0}^{\infty}|a_r|\,R^r\in[0,\infty].
\label{eq:majorant}
\end{equation}
Whenever $\widetilde\eta(\|f\|_W)<\infty$, one has
\begin{equation}
\|\eta\circ f \|_{W}\le \widetilde\eta(\|f \|_{W}).
\label{eq:compose0}
\end{equation}
\end{itemize}

\textit{Proof.}
We repeatedly use the trivial homogeneity $\|\alpha f\|_W=|\alpha|\,\|f\|_W$.

\vspace{0.1in}

\emph{(1) Subadditivity.}
By linearity of the Walsh transform, $\widehat{(f+g)}(S)=\widehat f(S)+\widehat g(S)$.
Thus
\begin{equation}
\begin{aligned}
\|f+g \|_{W}
&=\sum_{S}|\widehat f(S)+\widehat g(S)|\\
&\le \sum_{S}\big(|\widehat f(S)|+|\widehat g(S)|\big)
=\|f \|_{W}+\|g \|_{W}.    
\end{aligned}
\end{equation}


\emph{(2) Products and powers.}
Using the Walsh expansion $f(\sigma)=\sum_T \widehat f(T)\chi_T(\sigma)$ and $\chi_T\chi_U=\chi_{T\triangle U}$, one obtains the standard convolution identity
\begin{equation}
\widehat{(fg)}(S)=\sum_{T}\widehat f(T)\,\widehat g(S\triangle T),
\end{equation}
where $\triangle$ denotes symmetric difference.
Therefore,
\begin{equation}
\begin{aligned}
\|fg \|_{W}
&=\sum_{S}\Big|\sum_{T}\widehat f(T)\,\widehat g(S\triangle T)\Big|\\
&\le \sum_{S}\sum_{T}|\widehat f(T)|\,|\widehat g(S\triangle T)| \\
&=\sum_{T}|\widehat f(T)|\sum_{S}|\widehat g(S\triangle T)|\\
&=\Big(\sum_{T}|\widehat f(T)|\Big)\Big(\sum_{U}|\widehat g(U)|\Big)
=\|f \|_{W}\,\|g \|_{W},
\end{aligned}
\end{equation}
where we used that $S\mapsto S\triangle T$ is a bijection on subsets of $[N]$.
The power bound follows by iterating the product bound:
$\|f^{r} \|_{W}\le \|f^{r-1} \|_{W}\,\|f \|_{W}\le\cdots\le \|f \|_{W}^{r}$.

\vspace{0.1in}

\emph{(3) Affine functions.}
For $z(\sigma)=b+\sum_i w_i\sigma_i$, orthogonality of Walsh characters gives
$\widehat z(\varnothing)=b$, $\widehat z(\{i\})=w_i$, and $\widehat z(S)=0$ for $|S|\ge2$.
Hence $\|z \|_{W}=|b|+\sum_i|w_i|$.

\vspace{0.1in}

\emph{(4) Analytic composition.}
From the inversion formula $f(\sigma)=\sum_S \widehat f(S)\chi_S(\sigma)$ we have the pointwise bound
$|f(\sigma)|\le \sum_S|\widehat f(S)|=\|f\|_W$ for all $\sigma$.
Assuming $\widetilde\eta(\|f\|_W)<\infty$, the series $\sum_{r\ge0} a_r f(\sigma)^r$ converges absolutely for each $\sigma$ and defines $(\eta\circ f)(\sigma)$.
Using subadditivity and the power bound,
\begin{equation}
    \begin{aligned}
       \|\eta\circ f\|_W
&=\Big\|\sum_{r\ge0} a_r f^{r}\Big\|_W
\le \sum_{r\ge0} |a_r|\,\|f^{r}\|_W\\
&\le \sum_{r\ge0} |a_r|\,\|f\|_W^{r}
=\widetilde\eta(\|f\|_W), 
    \end{aligned}
\end{equation}

which proves Eq.~\eqref{eq:compose0}.
\hfill$\square$


\section{Theorem: A tame-majorant bound for additive feed-forward networks}

\textbf{Theorem.} \textit{(tame-majorant bound)}
Assume $\eta$ is analytic at the origin with absolute Taylor majorant $\widetilde\eta$ in Eq.~\eqref{eq:majorant}.
Define
\begin{equation}
\mathcal{W}\equiv
\max_{\ell,j}\sum_i |W_{ji}^{(\ell)}|,
\qquad
B\equiv
\max_{\ell,j} |b_j^{(\ell)}|,
\label{eq:SM_kappa_def}
\end{equation}
let $R_1\equiv 1$, and for $\ell=2,\dots,D$ define
\begin{equation}
R_\ell\equiv\widetilde\eta\big(B+\mathcal{W}R_{\ell-1}\big).
\label{eq:SM_R_recursion}
\end{equation}
Then the network output $g(\sigma)$ in Eq.~\eqref{eq:mlp} satisfies
\begin{equation}
\|g\|_W\le B+\mathcal{W}R_D.
\label{eq:SM_g0_bound}
\end{equation}

\textit{Proof.}
Each input coordinate is a Walsh character, so
\begin{equation}
\|u_i^{(1)}\|_W=\|\sigma_i\|_W=1=R_1,
\qquad i=1,\dots,N.
\label{eq:SM_input_norm}
\end{equation}
Now let $\ell\in\{2,\dots,D\}$ and assume $\|u_i^{(\ell-1)}\|_W\le R_{\ell-1}$ for all inputs to layer $\ell$.
Then homogeneity and subadditivity give
\begin{equation}
\begin{aligned}
    \|z_j^{(\ell)}\|_W
&=
\left\|
b_j^{(\ell)}+\sum_i W_{ji}^{(\ell)}u_i^{(\ell-1)}
\right\|_W\\
&\le
|b_j^{(\ell)}|+\sum_i |W_{ji}^{(\ell)}|\,\|u_i^{(\ell-1)}\|_W\\
&\le
B+\mathcal{W}R_{\ell-1},
\end{aligned}
\label{eq:SM_hidden_step}
\end{equation}
and hence, by the composition bound,
\begin{equation}
\|u_j^{(\ell)}\|_W
=
\|\eta\circ z_j^{(\ell)}\|_W
\le
\widetilde\eta\big(B+\mathcal{W}R_{\ell-1}\big)
=
R_\ell.
\label{eq:SM_hidden_activation}
\end{equation}
Induction therefore yields
\begin{equation}
\|u_j^{(D)}\|_W\le R_D,
\qquad j=1,\dots,w.
\label{eq:SM_last_hidden}
\end{equation}
Applying the same estimate to the formal output layer,
\begin{equation}
\|g\|_W
\le
B+\mathcal{W}R_D,
\end{equation}
which is Eq.~\eqref{eq:SM_g0_bound}.
\hfill$\square$

In the fully connected width-$w$ case with elementwise bound $|W_{ji}^{(\ell)}|\le s$, the first hidden layer has row sums at most $sN$, while all later hidden layers and the output layer have row sums at most $sw$.
Hence
\begin{equation}
\mathcal{W}\le s\max(N,w).
\label{eq:SM_fc_specialization}
\end{equation}

\section{Scaling of the majorant recursion and examples of $\widetilde\eta$}

Theorem~\eqref{eq:SM_g0_bound} reduces the expressibility question to the growth of the recursion~\eqref{eq:SM_R_recursion}.
We now record the resulting scaling for several common activations through their absolute Taylor majorants $\widetilde\eta$.

\textit{Polynomial activations.}
Let
\begin{equation}
\eta(x)=\sum_{r=0}^{p} a_r x^r
\end{equation}
be a degree-$p$ polynomial, and define
\begin{equation}
A_\eta\equiv\sum_{r=0}^{p}|a_r|.
\end{equation}
Then
\begin{equation}
\widetilde\eta(R)=\sum_{r=0}^{p}|a_r|R^r
\le
A_\eta(1+R)^p .
\label{eq:poly_majorant_simple}
\end{equation}
Set
\begin{equation}
Q_\ell\equiv 1+R_\ell .
\end{equation}
Since $R_1=1$, one has $Q_1=2$.
For $\ell=2,\dots,D$, Eqs.~\eqref{eq:SM_R_recursion} and \eqref{eq:poly_majorant_simple} imply
\begin{equation}
\begin{aligned}
Q_\ell
&=
1+\widetilde\eta\big(B+\mathcal{W}R_{\ell-1}\big) \\
&\le
1+A_\eta\big(1+B+\mathcal{W}R_{\ell-1}\big)^p \\
&\le
1+A_\eta\big(1+B+\mathcal{W}Q_{\ell-1}\big)^p \\
&\le
\alpha\,Q_{\ell-1}^{\,p},
\end{aligned}
\label{eq:poly_Q_recursion_compact}
\end{equation}
where
\begin{equation}
\alpha\equiv 1+A_\eta(1+B+\mathcal{W})^p .
\label{eq:poly_alpha}
\end{equation}
The last step uses $Q_{\ell-1}\ge1$, so
\begin{equation}
1+B+\mathcal{W}Q_{\ell-1}
\le
(1+B+\mathcal{W})Q_{\ell-1}.
\end{equation}

Iterating Eq.~\eqref{eq:poly_Q_recursion_compact} directly gives
\begin{equation}
Q_D
\le
2^{p^{D-1}}\,
\alpha^{\frac{p^{D-1}-1}{p-1}}.
\label{eq:poly_Q_explicit}
\end{equation}
Since
\begin{equation}
\alpha=1+A_\eta(1+B+\mathcal{W})^p
\le
C_\eta(2+B+\mathcal{W})^p,
\end{equation}
with $C_\eta = 1 + A_\eta$, and since $2+B+\mathcal{W}\ge 2$, it follows that
\begin{equation}
R_D\le Q_D
\le
(2+B+\mathcal{W})^{O(p^{D-1})}.
\end{equation}
Combining this with Theorem~\eqref{eq:SM_g0_bound} gives
\begin{equation}
\|g\|_W
\lesssim
(2+B+\mathcal{W})^{O(p^{D-1})}.
\label{eq:poly_g0_scaling}
\end{equation}
With $K\equiv 2+\mathcal{W}+B$, one may write
\begin{equation}
\|g\|_W\lesssim K^{O \left(p^{D-1}\right)},
\end{equation}
which is the form quoted in the main text.
In particular, if $K=\mathrm{poly}(N)$ and $D\le (1-\varepsilon)\log_p N$, then
\begin{equation}
\log\|g\|_W=o(N),
\qquad
\|g\|_W=\exp(o(N)).
\end{equation}

\textit{Entire nonpolynomial activations.}
For several standard entire functions the majorant is explicit:
\begin{equation}
\begin{gathered}
\eta(x)=e^x \ \implies\ \widetilde\eta(R)=e^R,\\
\eta(x)=\sin x \ \implies\ \widetilde\eta(R)=\sinh R,\\
\eta(x)=\cos x \ \implies\ \widetilde\eta(R)=\cosh R.
\end{gathered}
\end{equation}
In these cases $\widetilde\eta(R)$ grows as $\exp(\Theta(R))$ once $R$ is large, so the recursion~\eqref{eq:SM_R_recursion} can become very permissive whenever intermediate preactivations scale extensively with $N$.

\textit{Bounded analytic activations (finite radius of convergence).}
If $\eta$ is analytic at the origin but has complex singularities at finite distance, then $\widetilde\eta(R)$ diverges at a finite $R$, and Theorem~\eqref{eq:SM_g0_bound} yields a nontrivial ceiling only while the recursion stays below that divergence threshold.
A canonical example is $\eta(x)=\tanh x$, whose Taylor coefficients alternate in sign with the same magnitudes as those of $\tan x$.
Hence
\begin{equation}
\widetilde{\tanh}(R)=\tan R,
\end{equation}
which diverges at $R=\pi/2$.
Similarly, the logistic sigmoid can be written as
\begin{equation}
\sigma(x)=\tfrac12\bigl(1+\tanh(x/2)\bigr),
\end{equation}
so
\begin{equation}
\widetilde\sigma(R)\le \tfrac12\bigl(1+\tan(R/2)\bigr),
\end{equation}
which diverges at $R=\pi$.
In such bounded-activation settings, once optimization drives preactivations into saturation, the majorant recursion necessarily leaves its tame regime, and Eq.~\eqref{eq:SM_g0_bound} no longer provides a useful global upper bound.

\bibliography{main}

\end{document}